\documentclass[final,5p,times,twocolumn]{elsarticle}

\makeatletter
\def\ps@pprintTitle{%
 \let\@oddhead\@empty
 \let\@evenhead\@empty
 \def\@oddfoot{}%
 \let\@evenfoot\@oddfoot}
\makeatother

\usepackage{amssymb}
\usepackage{lineno}
\usepackage{color}
\usepackage{subfig}

\journal{Astroparticle Physics}

\begin{document}

\begin{frontmatter}



\title{Intrinsic neutron background of nuclear emulsions for directional Dark Matter searches}


\author[INFN-Na]{A. Alexandrov}
\author[Nagoya]{T. Asada}
\author[INFN-Na,Uni-Na]{A. Buonaura}
\author[GS]{L. Consiglio}
\author[GS]{N. D'Ambrosio}
\author[INFN-Na,Uni-Na]{G. De Lellis}
\author[INFN-Na]{A. Di Crescenzo}
\author[GS]{N. Di Marco}
\author[GS]{M. L. Di Vacri}
\author[Nagoya]{S. Furuya}
\author[INFN-Na,Uni-Na]{G. Galati}
\author[INFN-Na,Uni-Na]{V. Gentile}
\author[Nagoya]{T. Katsuragawa}
\author[GS]{M. Laubenstein}
\author[INFN-Na,Uni-Na]{A. Lauria}
\author[INFN-Rm,Uni-Rm]{P. F. Loverre}
\author[Nagoya]{S. Machii}
\author[INFN-Rm]{P. Monacelli}
\author[INFN-Na,Uni-Na]{M. C. Montesi}
\author[Nagoya]{T. Naka}
\author[INFN-Fr]{F. Pupilli\corref{cor1}}
\ead{fabio.pupilli@lnf.infn.it}
\author[INFN-Rm,Uni-Rm]{G. Rosa}
\author[Nagoya]{O. Sato}
\author[INFN-Na,Uni-Na]{P. Strolin}
\author[INFN-Na]{V. Tioukov}
\author[Nagoya]{A. Umemoto}
\author[Nagoya]{M. Yoshimoto}

 \address[INFN-Na]{INFN Sezione di Napoli, I-80125 Napoli, Italy}
 \address[Nagoya]{Nagoya University, J-464-8602 Nagoya, Japan}
 \address[Uni-Na]{Dipartimento di Fisica dell’Universit\`a Federico II di Napoli, I-80125 Napoli, Italy}
 \address[GS]{INFN - Laboratori Nazionali del Gran Sasso, I-67010 Assergi (L’Aquila), Italy}
 \address[INFN-Rm]{INFN Sezione di Roma, I-00185 Roma, Italy}
 \address[Uni-Rm]{Dipartimento di Fisica dell’Universit\`a di Roma Sapienza, I-00185 Roma, Italy}
 \address[INFN-Fr]{INFN - Laboratori Nazionali di Frascati dell’INFN, I-00044 Frascati (Roma), Italy}

\cortext[cor1]{Corresponding author}



\begin{abstract}
Recent developments of the nuclear emulsion technology led to the production of films with nanometric silver halide grains suitable
to track low energy nuclear recoils with submicrometric length. 
This improvement opens the way to a directional Dark Matter detection, thus
providing an innovative and complementary approach to the on-going WIMP searches.
An important background source for these searches is represented by neutron-induced nuclear recoils that can mimic
the WIMP signal. In this paper we provide an estimation of the contribution to this background from the intrinsic radioactive
contamination of nuclear emulsions. We also report the neutron-induced background as a function
of the read-out thre\-shold, by using a GEANT4 simulation of the nuclear emulsion,
showing that it amounts to about 0.06 per year per kilogram, fully compatible with the design of a 10 kg~$\times$~year exposure.
\end{abstract}


\end{frontmatter}

\section{Introduction}
\label{}

There is a compelling evidence \citep{evidence}, mostly coming from astrophysical observations \citep{rotation}, for the existence of Dark Matter
in the Universe.
Many direct search experiments are testing the WIMP (Weakly Interactive Massive Particle) hypothesis as Dark Matter candidate by measuring
the energy transferred to target nuclei through the WIMP-nucleus scattering. They look
for an excess of events over the expected background or for annually modulated signals \citep{annual_mod}
but they lead to controversial results: 
the annually modulated signal observed by DAMA/LIBRA \citep{DAMA} is in tension with upper limits on the interaction
cross section reported by other experiments \citep{LUX,XENON,CRESST,CDMS,PandaX}.

An unambiguous signature of
WIMPs would come from the observation of a non-isotropic signal with respect to the iso\-tro\-pically
distributed background: indeed, due to the relative motion of the solar system with respect to the
galactic halo, the WIMP flux should mostly come from the direction of this motion (i.e. from the
Cygnus constellation) and nuclear recoils induced by WIMPs should show a peak in the opposite
direction \citep{Spergel}.
Recoil tracks in solid targets are expected to be in the sub-$\mu$m range and require an unprecedented spatial accuracy. Therefore
many efforts have been pursued exploiting low-pressure gas detectors, like DRIFT \citep{DRIFT}, NEWAGE \citep{NEWAGE}, DMTPC \citep{DMTPC},
MIMAC \citep{MIMAC}; nevertheless these approaches require
large target volumes in order to be sensitive to low WIMP spin-independent interaction cross sections, and the small mass exposures performed so far are
limited by non-trivial scalability problems.

A recent proposal \citep{NEWS,NEWS1} foresees the use of nuclear emulsions both as target and as tracking device for a WIMP directional search.
Nuclear emulsion consists of a large number of sensitive silver bromide crystals, usually with linear dimensions of about 200 nm, like the one
used for OPERA films \citep{OPERAfilm}, dispersed in an organic gelatin matrix, composed essentially by hydrogen, carbon, nitrogen and oxygen.
Through-going ionizing particles release energy to the crystals creating sensitised sites along
their trajectories, the latent image. By the action of a mild chemical reducing agent, a
photographic developer, the conversion of halide to metallic silver selectively affects sensitised crystals.
After the development, followed by fixing and washing, the gelatin becomes optically transparent.
Three-dimensional paths of charged particles are finally visible as trails of silver grains using
optical microscopes.

The production of films with a crystal diameter
of 40$\pm$9 nm and a linear density of 11 crystals/$\mu$m was achieved for the first time in 2007 \citep{NIT}.
Further upgrades \citep{UNIT} have produced samples with a linear dimension of 
18$\pm$5 nm, one order of magnitude smaller than OPERA films \citep{OPERAfilm}; their li\-near density is $\sim$~29~crystals/$\mu$m, 
implying an intrinsic detector threshold of about 50 nm, if we consider tracks of at least two grains. This threshold corresponds to a
recoil energy of carbon (silver) nuclei of about 20 keV (110 KeV).

Radiogenic backgrounds have to be reduced and taken under control when dealing with signals as rare as the WIMP-nucleon scattering. 

The energy deposition per unit path length of WIMP-induced recoils is expected to be, for light and heavy nuclei respectively, one or two orders of
magnitude larger than the one due to electrons \citep{bkg_rejection};
therefore $\gamma$-rays and $\beta$-particles can be rejected by properly controlling the emulsion response, in terms of number of sensitised crystals
per unit path length (\textit{sensitivity}), through the chemical treatment of the emulsion itself.

$\alpha$-particles, essentially coming
from Uranium and Thorium decay chains, have energies of the order of MeV and can therefore be identified by 
their 3D range and rejected by an upper cut on the track length. For this purpose, a tomographic emulsion scanning with micrometric resolution, like the one performed in OPERA \citep{OPERAscanning} and able to reconstruct also per\-pen\-di\-cu\-lar tracks, will be
performed.

The most important background is represented by neutrons since they induce nuclear recoils with track lengths comparable to the WIMP-induced
ones.
For external sources we plan to inherit and adapt passive
shielding strategies successfully implemented in other dark matter experiments
and to operate the detector underground, for the reduction of the cosmogenic contribution.
Intrinsic contamination of nuclear emulsions by Uranium and Thorium traces is responsible for an irreducible neutron yield
through ($\alpha$,n) reactions and $^{238}$U spontaneous fission.
This neutron source is unavoidable
and can be minimized only by a proper selection and/or pre-treatment of the
emulsion components. The aim of this publication is to set the scale
of this irreducible background contribution.

After a description of the constituents and of the elemental composition
of nuclear emulsions we report the measurements of the intrinsic radioactive contamination. We then derive the related neutron
yield and its energy spectrum.
Based on a GEANT4 \citep{GEANT} simulation of nuclear emulsions, we estimate the intrinsic background induced by neutrons.
We finally conclude that,
even without a dedicated selection of the emulsion components, it is compatible with an emulsion exposure of $\mathcal{O}$(10)
kg~$\times$~year.

\section{Nuclear emulsion composition and intrinsic radioactive contamination}

Nuclear emulsions are composed by an organic gelatin matrix acting as retaining structure for their sensitive elements, the silver bromide crystals.
The emulsions developed for directional Dark Matter searches consist of a larger fraction of AgBr crystals with smaller dimension with respect
to the standard ones; crystals are also doped with iodine to increase their sensitivity to ionizing particles. It should be mentioned that the
presence of iodine slightly enhance the
detector sensitivity to spin-dependent WIMP interactions. Polyvinyl alcohol (PVA) is used to stabilise and
reduce the crystal growth. 
AgBr-I crystals are uniformly dispersed in a homogeneous mixture of gelatin and PVA.
The results reported hereafter are based on samples provided by
the Nitta Gelatin (cow's bones gelatin), the Kanto-Kagaku for AgBr and the Sigma-Aldrich for the PVA.
The mass fractions of the emulsion constituents are reported in \tablename~\ref{tab:constituents}, while in 
\tablename~\ref{tab:composition} the elemental composition is detailed. 
The emulsion composition, key ingredient for the neutron yield estimation and for the GEANT4 si\-mu\-lation, has been carefully determined
for light elements by
an elemental analyser (YANACO MT-6); their measurements were performed with a relative humidity ranging from 30\% to 40\% and have
an absolute uncertainty of 0.003. We have measured that a 10\% variation in relative humidity induces changes in H and O mass fractions
well within this uncertainty. 
The mass fraction of silver and bromine has been measured by an energy dispersive X-ray analysis with an absolute uncertainty of 0.02.
The density amounts to 3.43 g cm$^{-3}$.

\begin{table}[htpb]
\centering
\subfloat[Constituents of nuclear emulsion\label{tab:constituents}]{
\begin{tabular}{c|c}
\hline
\hline
Constituent & Mass Fraction \\
\hline
\hline
AgBr-I      & 0.78  \\

Gelatin     & 0.17 \\

PVA         & 0.05 \\
\hline
\hline
\end{tabular}
}

\subfloat[Elemental composition\label{tab:composition}]{
\begin{tabular}{c|c|c}
\hline
\hline
Element & Mass Fraction & Atomic Fraction \\
\hline
\hline
Ag      & 0.44        & 0.10 \\

Br      & 0.32        & 0.10 \\

I       & 0.019       & 0.004 \\

C       & 0.101        & 0.214 \\

O       & 0.074        & 0.118 \\

N       & 0.027        & 0.049 \\

H       & 0.016         & 0.410 \\

S       & 0.003        & 0.003 \\

\hline
\hline
\end{tabular}
}

\caption{Composition of the nuclear emulsion. The uncertainties on mass fractions are reported in the text.}

\end{table}

A sample of each component of the nuclear emulsion has been analysed by the Chemistry Service in Laboratori Na\-zio\-na\-li del Gran Sasso (LNGS, Italy),
with the Inductively Coupled Plasma Mass Spectrometry (ICP-MS) technique \citep{ICPMS}, in order to determine the Uranium and Thorium contamination; 
the instrument used for the analysis is a 7500a model from Agi\-lent Technologies and the results obtained have an uncertainty of 30\%.
The measured contaminations are reported in \tablename~\ref{tab:contamination} for all the constituents, together with the corresponding
acti\-vities obtained through the conversion factors 1 Bq kg$^{-1}$ ($^{238}$U) $\equiv$ 81$\times$10$^{-9}$ g g$^{-1}$ ($^{238}$U), 1 Bq kg$^{-1}$ ($^{232}$Th) 
$\equiv$ 246$\times$10$^{-9}$ g g$^{-1}$ ($^{232}$Th) \citep{conversion}. The upper limits on PVA are evaluated at 95\% CL.

\begin{table*}[t]
\begin{center}
\begin{tabular}{c|c|c}
\hline
\hline
Nuclide         & Contamination [10$^{-9}$ g g$^{-1}$] & Activity [mBq kg$^{-1}$] \\
\hline
\hline
\multicolumn{3}{c}{AgBr-I} \\
\hline
$^{232}$Th 	& 1.0 		      & 4.1                 \\
$^{238}$U	& 1.5		      & 18.5		  \\
\hline
\hline
\multicolumn{3}{c}{Gelatin} \\
\hline
$^{232}$Th 	& 2.7 		      & 11.0                \\
$^{238}$U	& 3.9		      & 48.1		  \\
\hline
\hline
\multicolumn{3}{c}{PVA} \\
\hline
$^{232}$Th 	& $< 0.5$             & $< 2.0$             \\
$^{238}$U	& $< 0.7$             & $< 8.6$		  \\
\hline
\hline
\end{tabular}
\end{center}
\caption{Results obtained by ICP-MS in terms of contamination and activity for the different constituents of the nuclear emulsion.
The estimated uncertainty is 30\%. The upper limits on PVA are evaluated at 95\% CL.}
\label{tab:contamination}
\end{table*}

\begin{table*}[htbp]
\begin{center}
\begin{tabular}{c|c|c|c}
\hline
\hline
Decay chain & Nuclide         & Contamination [10$^{-9}$ g g$^{-1}$] & Activity [mBq kg$^{-1}$] \\
\hline
\hline
\multicolumn{4}{c}{AgBr-I} \\
\hline
$^{232}$Th 	&  $^{228}$Ra     & $< 2.9$  		      & $< 12$                 \\
		&  $^{228}$Th     & $< 1.4$  		      & $< 5.5$  	  \\
\hline
$^{238}$U 	&  $^{226}$Ra     & $< 0.7$  		      & $< 8.9$                  \\
		&  $^{234}$Th     & $< 18$  		      & $< 220$ 		  \\
\hline
\hline
\multicolumn{4}{c}{Gelatin} \\
\hline
$^{232}$Th 	&  $^{228}$Ra     & $< 0.3$  		      & $< 1.3$                 \\
		&  $^{228}$Th     & $5.0 \pm 0.4$	      & $20 \pm 2$  	  \\
\hline
$^{238}$U 	&  $^{226}$Ra     & $0.19 \pm 0.05$	      & $2.4 \pm 0.6$                \\
		&  $^{234}$Th     & $< 6.4$  		      & $< 79$ 		  \\
\hline
\hline
\end{tabular}
\end{center}
\caption{Results obtained by $\gamma$-spectroscopy in terms of contamination and activity for the different constituents of the nuclear emulsion.
The upper limits are evaluated at 90\% CL.}
\label{tab:contamination_gamma}
\end{table*}

By weighting the measured activity of each constituent for its mass fraction, the total activity of nuclear emulsion can be calculated; for
PVA we assume conservatively a 95\% CL upper limit. The $^{238}$U activity amounts to 23$\pm$7~mBq~kg$^{-1}$
(i.e.~(1.9$\pm$0.6)$\times$10$^{-9}$ g g$^{-1}$), while the $^{232}$Th one is 5.1$\pm$1.5 mBq~kg$^{-1}$ (i.e.~(1.3$\pm$0.4)$\times$10$^{-9}$ g g$^{-1}$).
The reported errors are dominated by the 30\% uncertainty in the radioactive con\-ta\-mi\-na\-tion measurements.
By assuming a null contribution from PVA, the previous contaminations are reduced by $\sim$ 2\%.

In nature Uranium and Thorium decay chains are found in se\-cu\-lar equilibrium; 
therefore, in order to calculate the neutron
yield from ($\alpha$,n) reactions, the same activity of the parent is assumed for all
the daughter nuclides in the chain.
Nevertheless, the human intervention during the manufacturing process may alter the equilibrium by artificially modifying the quantity
of some nuclides in the chain. In order to verify the validity of this assumption and also to cross check the measured activities,
the samples were analysed with the $\gamma$-spectroscopy technique sensitive to $\gamma$-active daughter nuclides.
The measurements have been performed in the low background facility STELLA (SubTErranean Low Level Assay) of the LNGS \citep{STELLA}
with germanium detectors; a sample
of about 300 g of AgBr-I and one of about 500 g of gelatin were used and the data taking lasted for about 2 and 3 weeks respectively.

As reported in \tablename~\ref{tab:contamination_gamma}, the results for the AgBr-I are in fair agreement with the mass spectrometry measurement
and there is no indication for broken decay chains.
For the gelatin, the measurements gave comparable results for the $^{232}$Th chain, while the $\gamma$-spectroscopy measured
a concentration of $^{226}$Ra in the $^{238}$U chain about 20 times smaller than the parent isotope, with a measured value of
2.4$\pm$0.6~mBq~kg$^{-1}$. This measurement suggests a break in the secular equilibrium of the decay chain at
this point. For the gelatin, secular equilibrium is assumed for the upper part of the $^{238}$U chain, using the
activity measured by mass spectrometry, while, for the lower part, nuclides will be considered in equi\-li\-brium
with $^{226}$Ra and the activity measured with $\gamma$-spectroscopy will be used. Therefore the nuclear emulsion activity for nuclides of the
$^{226}$Ra sub-chain, including also the activities of AgBr-I and PVA weighted by the corresponding mass fraction,
is 15$\pm$5~mBq~kg$^{-1}$.

\section{Radiogenic neutron yield estimation}
The total emulsion activities derived in the previous section are essential to determine the intrinsic neutron yield; indeed
Uranium and Thorium decay chains are responsible of neutron generation from detector materials
through spontaneous fission and ($\alpha$,n) interactions.

\begin{table*}[t]
\begin{center}
\begin{tabular}{c|c|c}
\hline
\hline
Process                               & SOURCES simulation 	& Semi-analytical calculation	\\
				      & [kg$^{-1}$ y$^{-1}$]      & [kg$^{-1}$ y$^{-1}$] \\
\hline
\hline
($\alpha$, n) from $^{232}$Th chain   & 0.12$\pm$0.04		    & 0.11$\pm$0.03	        \\
($\alpha$, n) from $^{238}$U chain    & 0.27$\pm$0.09		    & 0.26$\pm$0.08	        \\
Spontaneous fission     	      & 0.8$\pm$0.3		    & 0.8$\pm$0.3		\\
\hline
\hline
Total flux               	      & 1.2$\pm$0.4		    & 1.2$\pm$0.4		\\
\hline
\hline
\end{tabular}
\end{center}
\caption{Neutrons per kilogram per year due to ($\alpha$, n) and spontaneous fission reactions in the
  nuclear emulsion, evaluated with the SOURCES code and semi-analytical calculation.}
\label{tab:neutronbkg}
\end{table*}

Many nuclides undergo spontaneous fission, but the only re\-le\-vant contribution comes from $^{238}$U, since the number of fissions per decay of
other natural elements is at least two orders of magnitude lower. The neutron production rate R is calculated as:

\begin{equation}
 R_{sf} = A \times \psi \times n
\end{equation}
where A is the $^{238}$U trace activity, $\psi=5.45 \times 10^{-7}$ \citep{sf} is the fission probability per decay and $n=2.07$ \citep{sf}
is the average number of neutrons emitted per fission.

The rate from ($\alpha$,n) reactions depends on the rate and energy of $\alpha$ decays and on the materials in which the radioactive
contaminations are embedded. It can be determined with a semi-analytical approach exploiting the experimental neutron yield from
different elements and the method reported in reference \citep{Heaton}; this approach relies on the conservative assumption that the nuclear
emulsion film is thick enough to let all $\alpha$ particles to stop or interact inside; if this condition is not satisfied, the calculated
neutron flux has to be considered as an overestimation.
It is also assumed that ($\alpha$,n) reactions on high Z elements, such as Ag and Br, do not contribute to the total neutron flux;
indeed for these elements the Coulomb barrier of electrons surrounding the nucleus is of the same order of the highest $\alpha$ energies
from natural radioactive chains, and therefore the cross section of the reaction is highly suppressed. The neutron rate is given by:

\begin{equation}\label{eq:eq_n}
 R_{(\alpha,n)} = \sum_{i} B_{i} \times y^{c}_{i}
\end{equation}
where $i$ runs over the $\alpha$-emitting nuclides in the Uranium and Thorium chains;
$B_{i}$ is the total emulsion activity for the $i$-th nuclide, as determined in the previous section; $y^{c}_{i}$ is the neutron
yield of the whole nuclear emulsion, that can be written in terms of the individual elements as:

\begin{equation}\label{eq:eq_yield}
 y^{c}_{i} = \sum_{j} \frac{w_{j} S^{m}_{j}(E_{i})}{S^{m}_{c}(E_{i})} y_{i,j}(E_{i}) \zeta_{i}
\end{equation}
where $j$ runs over all the elements of the nuclear emulsion; $w_{j}$ is the mass fraction of the $j$-th element
(\tablename~\ref{tab:composition}); $y_{i,j}(E_{i})$ is the neutron yield of the $j$-th element for an $\alpha$ particle at ener\-gy $E_{i}$ 
emitted by the $i$-th nuclide, as derived from \citep{Heaton}; $\zeta_{i}$ is the branching ratio of the $i$-th nuclide decay.
$S^{m}_{c}(E_{i})$ is the total mass stopping power of the nuclear emulsion at the given ener\-gy and, assuming the Bragg additivity
rule, it is evaluated as the weighted sum of the mass stopping powers of its elements:

\begin{equation}
 S^{m}_{c}(E)= \sum_{j} w_{j} S^{m}_{j}(E)
\end{equation}
where $S^{m}_{j}$ is given by the stopping power divided by the density $\rho_{j}$ of the $j$-th element .

The mass stopping powers of each element were taken from the ASTAR database \citep{astar}; since bromine is not included in the database,
the stopping power values were calculated through the Bethe's formula, using for the mean excitation potential the following empirical formula
\citep{tsoulfanidis}:

\begin{equation}
 I(eV) = (9.76 + 58.8Z^{-1.19})Z
\end{equation}
with $Z=35$. For the bromine density the value 3.1028 g cm$^{-3}$ was used.

In order to cross check the estimation obtained with this approach and to calculate also the neutron spectrum, the SOURCES code \citep{sources}
was used. The inputs used are the atomic fractions reported in \tablename~\ref{tab:composition}, the number of
atoms for each $\alpha$-emitting nuclide, derived from the activities reported in the previous section, and the atomic
fractions of target nuclides, accounting for the natural isotopic abundances \citep{isotope} of each ele\-ment.

In the calculation, $\alpha$-emitters are assumed to be uniformly distributed  in the material. Furthermore, as for the semi-analytical calculation,
the thick-target assumption is used and high Z elements are not considered as target nuclides.

The original version of SOURCES describes ($\alpha$,n) reactions only up to 6.5 MeV $\alpha$-energies. This would limit the
reliability of the results, both in terms of estimated neutron flux and spectrum, since the reaction cross section and the average neutron
energy both increase with the incident $\alpha$ energy. To overcome this limitation, a modified version of SOURCES \citep{sources_mod},
able to deal with $\alpha$-energies up to 10 MeV, was used.

The neutron production rates from spontaneous fission and ($\alpha$,n) reactions, as determined with both the semi-analytical calculation
and the SOURCES code, are reported in \tablename~\ref{tab:neutronbkg}.
Following the approach reported in reference~\citep{Heaton}, we have estimated an uncertainty on $y_{c}$ (Eq.~\ref{eq:eq_yield})
of about 9\% due to the mass stopping powers, the emulsion composition and the neutron yields of each element. The total uncertainty on the
neutron production rate (Eq.~\ref{eq:eq_n}) is therefore dominated by the 30\% contribution due to the activity of $\alpha$-decaying nuclides.
The SOURCES code uncertainty of 17\% quoted in reference~\citep{sources}, convoluted with the 30\% uncertainty in the activity estimation, gives
a total uncertainty for this computation of 34\%.

The two approaches give comparable results and the flux due to the intrinsic radioactive contamination is expected to be of
the order of 1 neutron per year per kilogram of nuclear emulsion. 
It is worth noticing that the effect of the reduced activity measured with the $\gamma$-spectroscopy for the $^{226}$Ra sub-chain of gelatin amounts to
less than 10\% in the total neutron flux per year per kilogram.

\begin{figure}[htbp]
\begin{center}
\includegraphics[width=9 cm]{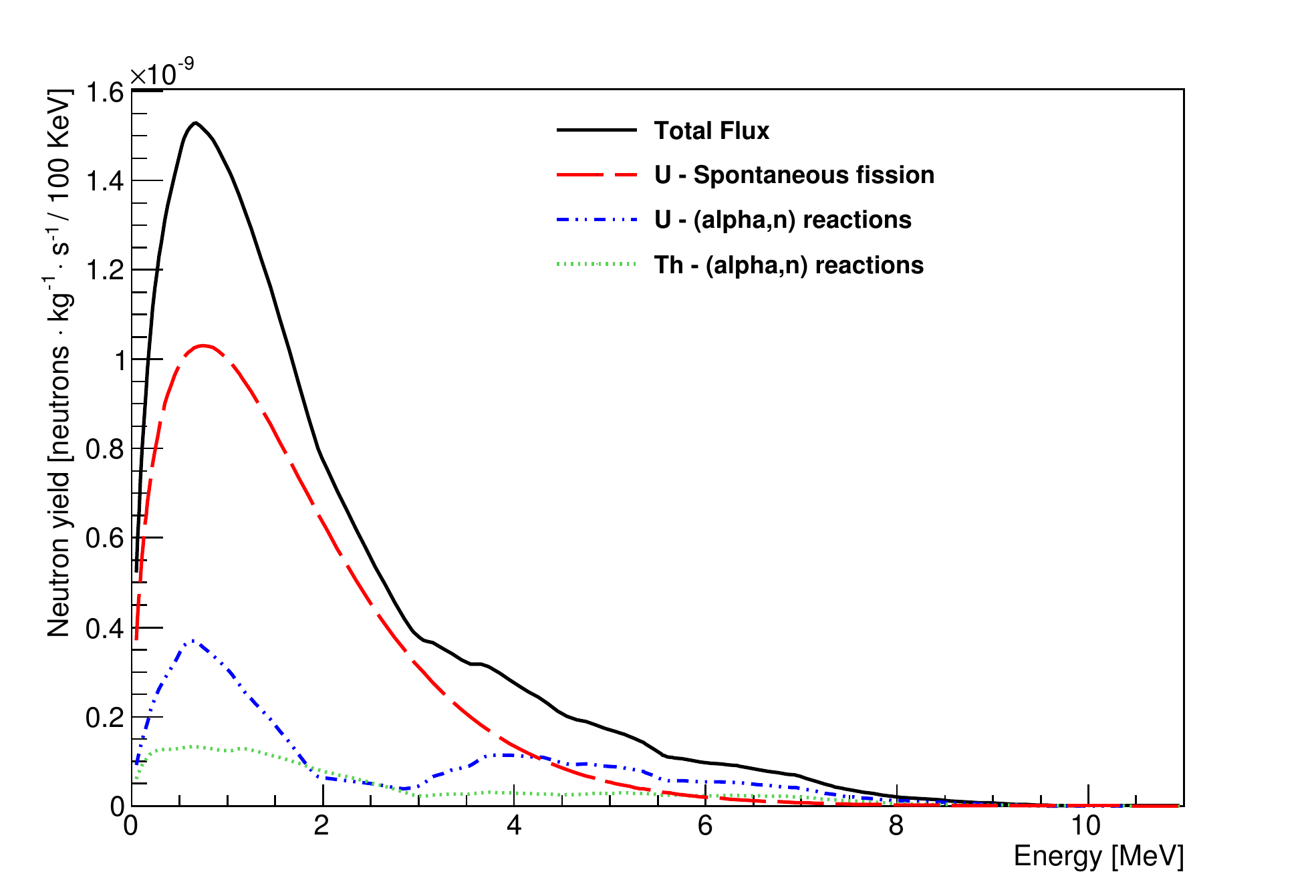}
\end{center}
\caption{Total neutron energy spectrum (solid black line); in dashed red the contribution from $^{238}$U spontaneous fission is shown, 
while in dot-dashed blue and dotted green the contributions from ($\alpha$,n) reactions due to nuclides in the $^{238}$U and $^{232}$Th
chains respectively are displayed.}
\label{fig:neutronspectrum}
\end{figure}

The energy spectrum of the produced neutrons, as calculated with SOURCES, is reported in \figurename~\ref{fig:neutronspectrum};
it it is peaked at 0.7 MeV, with a mean value of about 2 MeV.

\section{Background estimation} 

\begin{figure}[htbp]
\begin{center}
\includegraphics[width=8 cm]{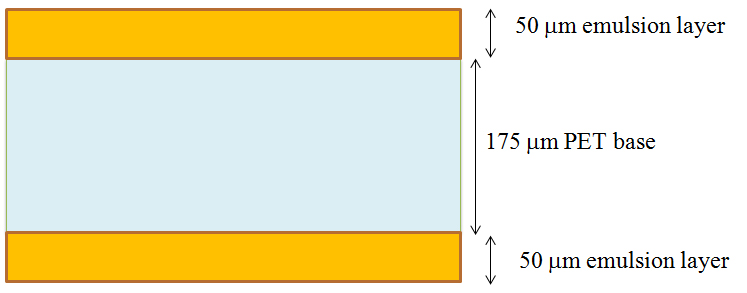}
\end{center}
\caption{Double coated emulsion film.}
\label{fig:emulsion_film}
\end{figure}

In order to estimate the detectable background due to radiogenic neutrons produced by the intrinsic radioactive contamination of the nuclear emulsions,
a GEANT4 \citep{GEANT} based si\-mu\-lation was performed. 
In the simulated setup two 50 $\mu$m thick nuclear emulsion layers are coated on both sides of a 175 $\mu$m thick plastic base
made of Polyethylene terephthalate (PET). This double coated emulsion film is sketched in \figurename~\ref{fig:emulsion_film}.
With the ICP-MS technique, an upper limit of 20$\times$10$^{-12}$ g g$^{-1}$ at 95\% CL for both
Uranium and Thorium was measured for the PET intrinsic contamination. The resulting upper limit on the neutron flux, estimated with the
same approach explained in the previous section, is 60 times lower than
the contribution from nuclear emulsion, and therefore is neglected.

1 kg of nuclear emulsion with the same chemical composition
described in \tablename~\ref{tab:composition} has been simulated.

The fraction of interacting neutrons depends on the ratio between the volume and the
surface of nuclear emulsion. For a quantitative estimate,
in this work we assume that 50 double coated emulsion films, with a surface of of 25$\times$25 cm$^{2}$, are piled-up for a total thickness of
13.75 mm, corresponding to an emulsion thickness of 5 mm.

One million neutrons were generated with an isotropic angular distribution; they were uniformly distributed in the emulsion material only,
given the negligible contribution of the plastic base to the neutron flux.
The ener\-gy spectrum was generated according to \figurename~\ref{fig:neutronspectrum}. The relevant processes in this energy range are the
elastic scattering, the inelastic scattering and the neutron capture.
73.8\% of neutrons undergoes at least one interaction, while 36.8\% are involved also in secondary interactions.
About 65\% of neutron interactions happen in PET.
\tablename~\ref{tab:process} shows the fractions of different interaction processes, while
the fractions of recoiled particles are reported in \tablename~\ref{tab:recoils}.

\begin{table}[htbp]
\begin{center}
\begin{tabular}{c | c}
\hline
\hline
  Process &  Fraction \\
 \hline
 \hline
 elastic scattering  &   0.973 \\
 inelastic scattering     & 0.023 \\
  neutron-capture   &  0.004 \\
 \hline
 \hline
 \end{tabular}
 \end{center}
 \caption{Interaction processes and their fractions.}
 \label{tab:process}
 \end{table}

\begin{table}[htbp]
\begin{center}
\begin{tabular}{c | c}
\hline
\hline
  Recoils &  Fraction \\
 \hline
 \hline
 nuclei  &   0.394 \\
 protons      & 0.605 \\
  $\alpha$-particles   &  0.001 \\
 \hline
 \hline
 \end{tabular}
 \end{center}
 \caption{Recoiled particles and their fractions.}
 \label{tab:recoils}
 \end{table}

\begin{figure}[htbp]
\begin{center}
\begin{minipage}{14pc}
\includegraphics[scale=0.37]{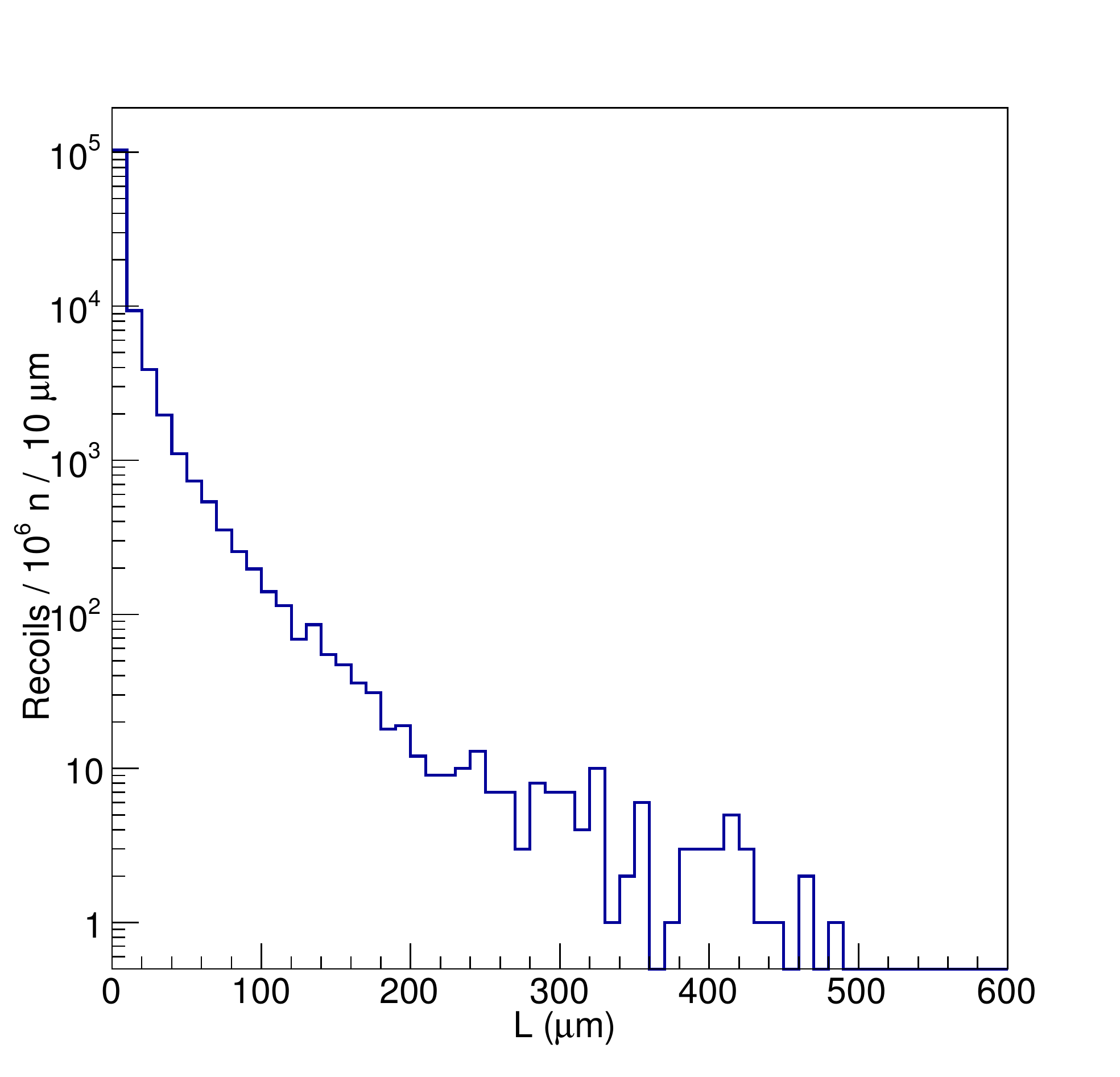}
\end{minipage}
\begin{minipage}{14pc}
\includegraphics[scale=0.37]{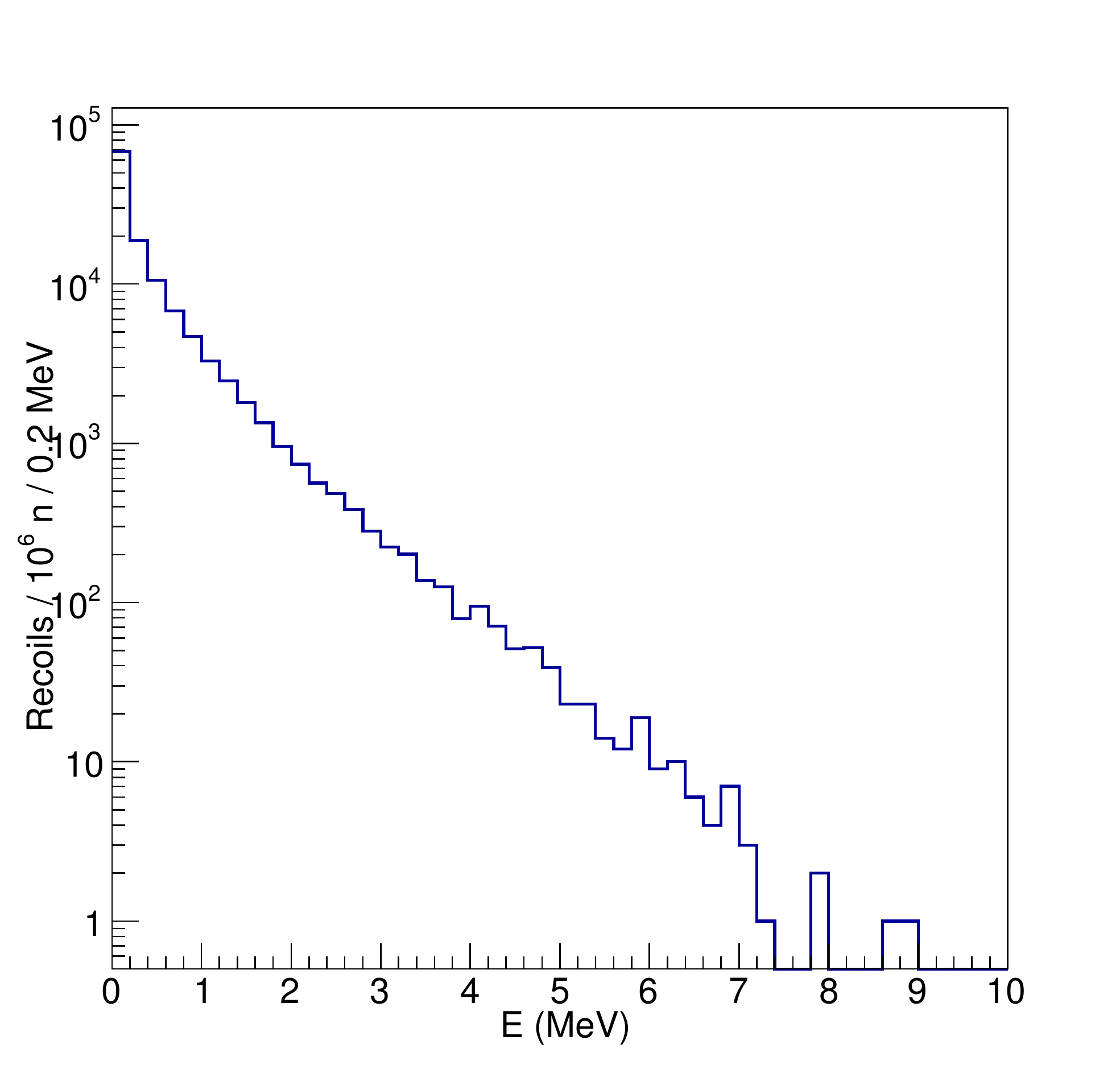}
\end{minipage}
\end{center}
\caption{Visible track length (top) and energy spectrum (bottom) for proton recoils.}
\label{fig:proton_recoil}
\end{figure}

In order to estimate the background contribution to the proposed Dark Matter search, we have evaluated the track length
of the induced recoils. \figurename~\ref{fig:proton_recoil} reports the visible track length distribution of proton recoils, 
defined as the distance between the first and last hit when they both happen in the active emulsion layers.
This length ranges from a few nm to several hundreds
$\mu$m. The proton ener\-gy distribution is also displayed.

\begin{figure}[htbp]
\begin{center}
\begin{minipage}{14pc}
\includegraphics[scale=0.37]{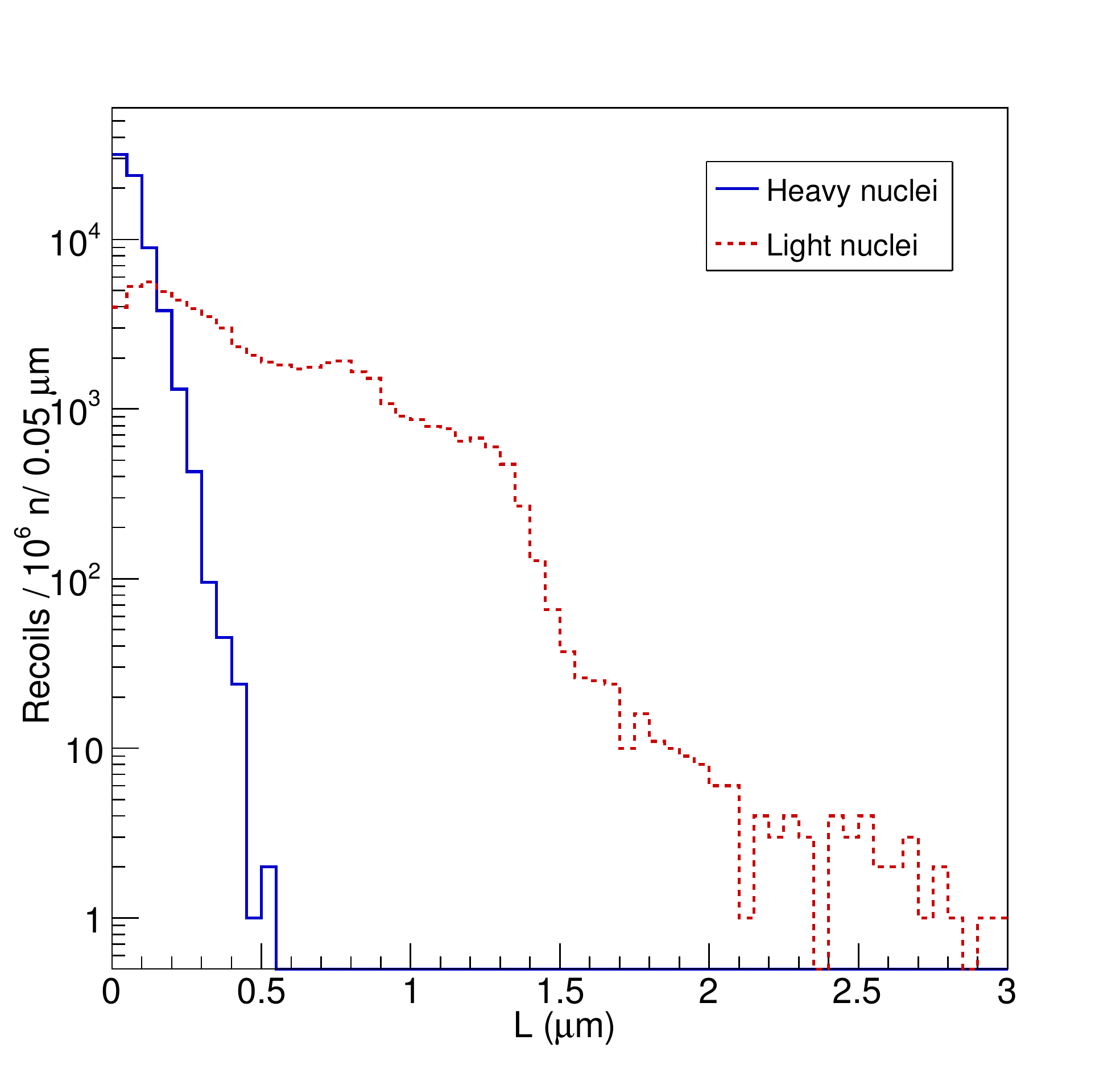}
\end{minipage}
\begin{minipage}{14pc}
\includegraphics[scale=0.37]{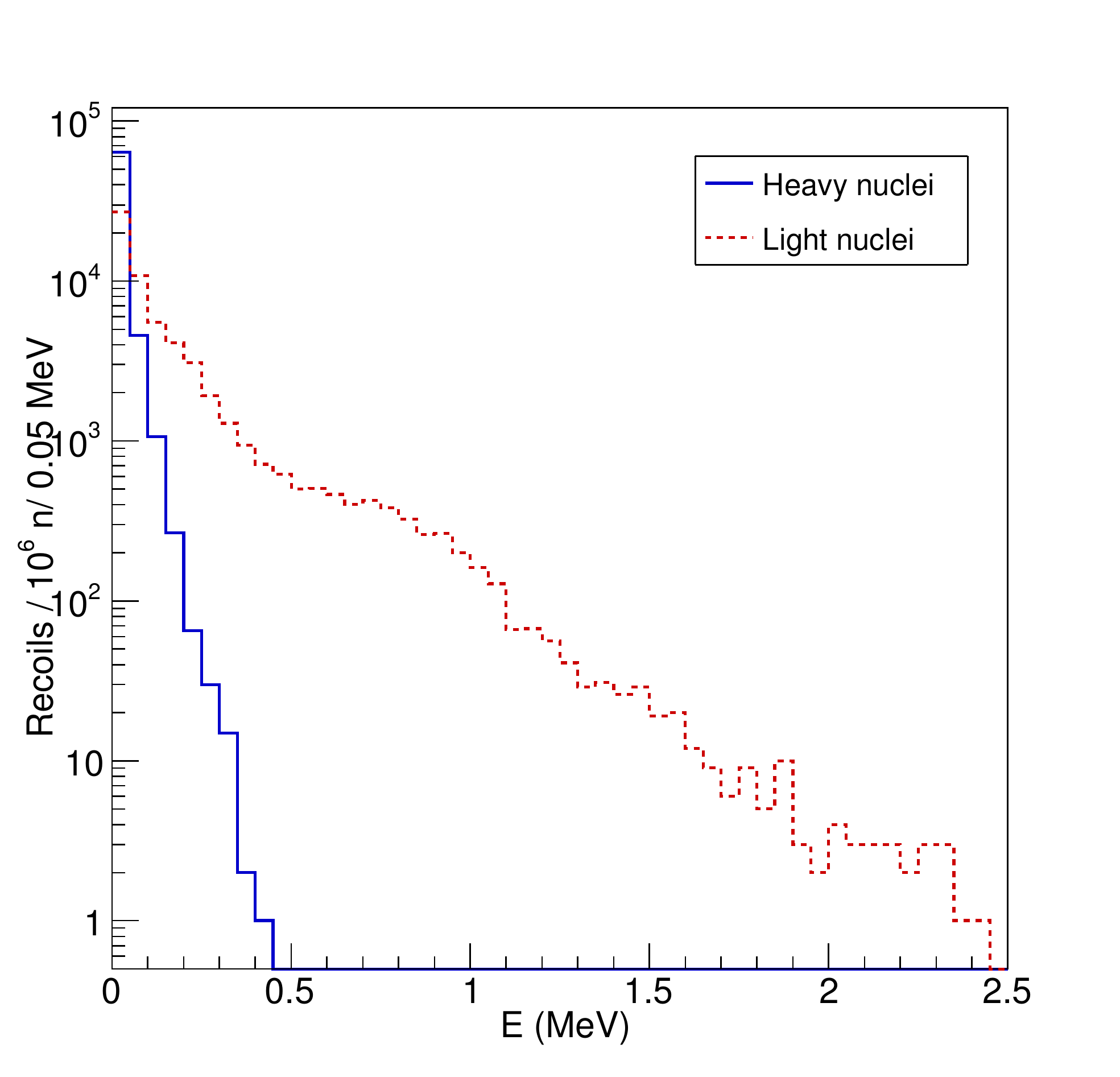}
\end{minipage}
\end{center}
\caption{Visible track length (top) and energy spectrum (bottom) for heavy nuclei (Ag,Br,I) (solid blue line) and light nuclei (C,N,O,S) (dashed red line).}
\label{fig:nuclear_recoil}
\end{figure}

Nuclear recoils have a softer spectrum and therefore they exhibit a shorter visible track length, not exceeding 3 $\mu$m for
light nuclei (C,N,O,S) and 1 $\mu$m for heavy nuclei (Ag,Br,I), as shown in \figurename~\ref{fig:nuclear_recoil}.

\begin{table*}[htbp]
\begin{center}
\begin{tabular}{c|p{1cm}|p{1cm}|c}
\hline
\hline
            & \multicolumn{3}{c}{WIMP mass (GeV)} \\
            & \centering 10     &  \centering  100     & 1000 \\
\hline
\hline
Recoil nucleus &  \multicolumn{3}{c}{Maximum recoil length (nm)} \\
\hline
C 	&  \centering 100     & \centering 350  & 430 \\
Ag	&  \centering 11      & \centering 140  & 480 \\
\hline
\hline
\end{tabular}
\end{center}
\caption{Maximum  WIMP-induced recoil length of Carbon and Silver nuclei for three different WIMP masses.}
\label{tab:maxrecoil}
\end{table*}

The threshold on detectable track lengths in nuclear emulsion depends on the read-out technology:
fully automated optical microscopes
have shown a good detection efficiency above 200 nm \citep{NEWS1}; on-going R\&D activities
envisage to lower the detectable track length down to the intrinsic detector threshold of about 50 nm.
We consider in this analysis different detector thresholds between 50 and 200 nm.
On the other side an upper limit of 1 $\mu$m on the track length can be introduced to suppress the background from radiogenic $\alpha$ particles,
since the signal is expected below 1 $\mu$m even for large ($\mathcal{O}$(TeV)) WIMP masses, as shown in \tablename~\ref{tab:maxrecoil}. 
This cut is effective in suppressing also the neutron-induced proton recoils.
The fractions of neutron-induced recoils below this cut, as a function of the read-out threshold,
are reported in \tablename~\ref{tab:recoils1}: only 7\% to 14\% of the intrinsic neutron flux contributes to the background.

\begin{table}[htbp]
\begin{center}
\begin{tabular}{c | c }
\hline
\hline
  Threshold [nm] &  Fraction \\
  \hline
  \hline
 50  & 0.138   \\
 100 & 0.104   \\
 150 & 0.086   \\
 200 & 0.073   \\
 \hline
 \hline
 \end{tabular}
 \end{center}
 \caption{Fraction of detectable neutron-induced recoils as a function of the read-out threshold, with an upper cut on the track length 
 of 1 $\mu$m.}
 \label{tab:recoils1}
 \end{table}

A further background suppression can be obtained by exploiting the directionality information: indeed, as previously stated, the signal produced
by WIMP-induced nuclear recoils is expected to have a directional signature.
We assume that the surface of the nuclear emulsion will be placed in parallel to the Sun direction in the Galaxy, thanks to an equatorial telescope.
In this re\-fe\-ren\-ce frame, the longitudinal component of the scattering angle is dominant on ave\-ra\-ge and its projection $\phi$ on the plane
of the emulsion surface can be defined.
A further reduction of $\sim$ 36\% of the neutron-induced background can be achieved with the cut $ -1 <\phi < 1$, accounting for the insensitivity
to the sense.
Under these assumptions, the intrinsic neutron-induced background would be 0.06 $\div$ 0.11 per year per kilogram.

\section{Summary}

An unambiguous proof of the existence of Dark Matter in the form of WIMP particles would come from the directional observation of induced nuclear
recoils. The use of nuclear emulsions both as target and as detector is promising.

One of the most important background for such an experiment comes from proton and nuclear recoils induced by neutrons from the intrinsic radioactive
contamination of nuclear emulsions. The trace activity of $^{238}$U and $^{232}$Th radioactive chains have been determined by measuring
each nuclear emulsion component by Inductively Coupled Plasma Mass Spectrometry (ICP-MS): the total emulsion activity is 23~mBq~kg$^{-1}$ for the Uranium
chain and 5.1~mBq~kg$^{-1}$ for the Thorium chain with an uncertainty of 30\%. 
The $\gamma$-spectroscopy analysis, performed in order to cross check the results of ICP-MS and to verify the secular equilibrium of the radioactive
chains, confirmed the results except for a lower activity of the $^{226}$Ra sub-chain of $^{238}$U for gelatin.

The measured radioactive contaminations have been used to calculate the
neutron flux generated by spontaneous fission of $^{238}$U and by ($\alpha$,n) reactions with the emulsion elements: it amounts to 1.2$\pm$0.4
neutrons per year per kilogram of nuclear emulsion and the energy spectrum has a mean value of about 2 MeV.

A GEANT4 simulation demonstrates that, under certain assumptions on the detector geometry and on the read-out stra\-te\-gy, the detectable 
neutron-induced background can be reduced down to 0.06 per year per kilogram.

Therefore the neutron-induced background due to the intrinsic radioactive contamination allows the design
of an emulsion detector with an exposure of about 10 kg~$\times$~year.
A careful selection of the emulsion components and a better control of their production 
could further increase the radiopurity, thus extending the detector exposure.


\section*{Acknowledgments}
We warmly thank C. Galbiati and P. Mosteiro for useful discussions on the measurements performed and on the estimate of the neutron
background. We are also indebted with V. Kudryavtsev for helps with the SOURCES code. We thank the LNGS chemistry service for the mass spectrometry
measurements.
We would like to thank K. Oyama of Chemical Instrumentation Facility, Nagoya University, Japan for the measurements of nuclear emulsion composition.
This work was also supported by JSPS KAKENHI Grant-in-Aid for Young Scientists (B) Number 25800140 and Grant-in-Aid for Scientific Research on
Innovative Areas Number 14429969.





\end{document}